\documentclass[twocolumn,aps,prl,preprintnumbers]{revtex4}
\usepackage{}
\usepackage{bbm}
\usepackage[latin9]{inputenc}
\usepackage{amsmath}
\usepackage{amssymb}
\usepackage{graphicx}
\usepackage{mathrsfs}
\usepackage{amsfonts}
\usepackage{amsthm}
\usepackage{color}
\usepackage{txfonts}
\usepackage{lipsum}
\usepackage[colorlinks=true,citecolor=blue,linkcolor=blue,urlcolor=blue,anchorcolor=blue]{hyperref}%
\hypersetup{colorlinks=true,citecolor=blue,linkcolor=blue,urlcolor=blue}
\providecommand{\U}[1]{\protect\rule{.1in}{.1in}}
\makeatletter

\newcommand{\Rmnum}[1]{\expandafter\@slowromancap\romannumeral #1@}
\makeatother
\makeatletter
\@ifundefined{textcolor}{}
{
\definecolor{BLACK}{gray}{0}
\definecolor{WHITE}{gray}{1}
\definecolor{RED}{rgb}{1,0,0}
\definecolor{GREEN}{rgb}{0,1,0}
\definecolor{BLUE}{rgb}{0,0,1}
\definecolor{CYAN}{cmyk}{1,0,0,0}
\definecolor{MAGENTA}{cmyk}{0,1,0,0}
\definecolor{YELLOW}{cmyk}{0,0,1,0}
}
\makeatother
\begin{document}
\title{Second-order topological insulator in breathing square lattice of magnetic vortices}
\author{Z.-X. Li$^{1}$}
\author{Yunshan Cao$^{1}$}
\author{X. R. Wang$^{2,3}$}
\author{Peng Yan$^{1}$}
\email[Corresponding author: ]{yan@uestc.edu.cn}
\affiliation{$^{1}$School of Electronic Science and Engineering and State Key Laboratory of Electronic Thin Films and Integrated Devices, University of Electronic Science and Technology of China, Chengdu 610054, China}
\affiliation{$^{2}$Physics Department, The Hong Kong University of Science and Technology,
 Clear Water Bay, Kowloon, Hong Kong}
\affiliation{$^{3}$HKUST Shenzhen Research Institute, Shenzhen 518057, China}
\begin{abstract}
We study the topological phase in dipolar-coupled two-dimensional breathing square lattice of magnetic vortices. By evaluating the quantized Chern number and $\mathbb{Z}_{4}$ Berry phase, we obtain the phase diagram and identify that the second-order topological corner states appear only when the ratio of alternating bond lengths goes beyond a critical value. Interestingly, we uncover three corner states at different frequencies ranging from sub GHz to tens of GHz by solving the generalized Thiele's equation, which has no counterpart in condensed matter system. We show that the emerging corner states are topologically protected by a generalized chiral symmetry of the quadripartite lattice, leading to particular robustness against disorder and defects. Full micromagnetic simulations confirm theoretical predictions with a great agreement. A vortex-based imaging device is designed as a demonstration of the real-world application of the second-order magnetic topological insulator. Our findings provide a route for realizing symmetry-protected multi-band corner states that are promising to achieve spintronic higher-order topological devices.
\end{abstract}

\maketitle
\section{INTRODUCTION}
Over the past few years, the concept of higher-order topological insulator (HOTI) \cite{Benalcazar2017,Bernevig2017,EzawaPRL2018,Song2017,Langbehn2017,Schindler2018,Queiroz2019} has attracted significant attention by the community for the peculiar symmetry-protected states emerging in device corners and hinges. HOTIs have been studied in the broad field of acoustics \cite{Xue2019,Ni2019,Yang2019,He2019,Zhang2019_1}, photonics \cite{Noh2018,Hassan2018,Mittal2018,Chen2019,Xie2019}, mechanics \cite{Serra-Garcia2018,Fan2019,Wakao2019}, electric circuits \cite{Imhof2018,Serra2019,EzawaPRB2018,Bao2019}, and recently, spintronics \cite{Linpj2019,Li2019,Sil2019}. As a new member of topological insulators (TIs) \cite{Hasan2010,Qi2011}, HOTIs go beyond the conventional bulk-boundary correspondence and are characterized by a few new topological invariants \cite{EzawaPRL2018,Song2017,Langbehn2017,Wan2017,Slager2015}. Very recently, it has been suggested that the $\mathbb{Z}_{N}$ Berry phase quantized to $2\pi/N$ is a useful tool to characterize the higher-order topological phase \cite{Wakao2019,Li2019,Zak1985,Kariyado2018,Hatsugai2011,Araki2019}. Since topologically protected corner states are robust against disorder and defects, they can act as localized oscillators. Previous work, however, focused on corner states only with a single frequency \cite{Xue2019,Ni2019, Noh2018,Hassan2018,Imhof2018,EzawaPRB2018}. It is intriguing to pursue topologically stable multi-mode corner phases for practical applications, such as imaging \cite{Zhang2019AM}.

Similar to other (quasi-)particles, the collective motion of magnetic textures, such as vortex \cite{Wachowiak2002,Waeyenberge2006}, bubble \cite{Makhfudz2012,Moon2014}, and skyrmion \cite{Binz2009,Jiang2015}, can also exhibit the behavior of waves \cite{Vogel2010,Jung2011,Han2013,Behncke2015,Schulte2016,Yang2017,Mruczkiewicz2016}. Notably, single band topological chiral edge states have been demonstrated in a two-dimensional honeycomb lattice of magnetic vortices (or bubbles) by solving the massless Thiele's equation that describes the coupled dynamics of magnetic solitons \cite{Kim2017}. We recently generalized the approach by including both a second-order inertial term and a third-order non-Newtonian gyroscopic term to interpret the emerging multiband nature of chiral edge states observed in honeycomb lattice of magnetic skyrmions \cite{Li2018PRB}. More recently, we have predicted the second-order TIs in breathing lattice of magnetic vortices \cite{Linpj2019,Li2019} and have shown that the generalized chiral symmetry protects the topological corner states. However, the high-frequency-band corner modes have not been well addressed. A prominent demonstration of the practical application of corner states is still lacking. In addition, the second-order topological phase in quadripartite lattice is yet to be reported, although the issue has received thorough investigations in both tripartite \cite{Linpj2019} and sexpartite \cite{Li2019} vortex lattices.

\begin{figure*}[ptbh]
\begin{centering}
\includegraphics[width=0.86\textwidth]{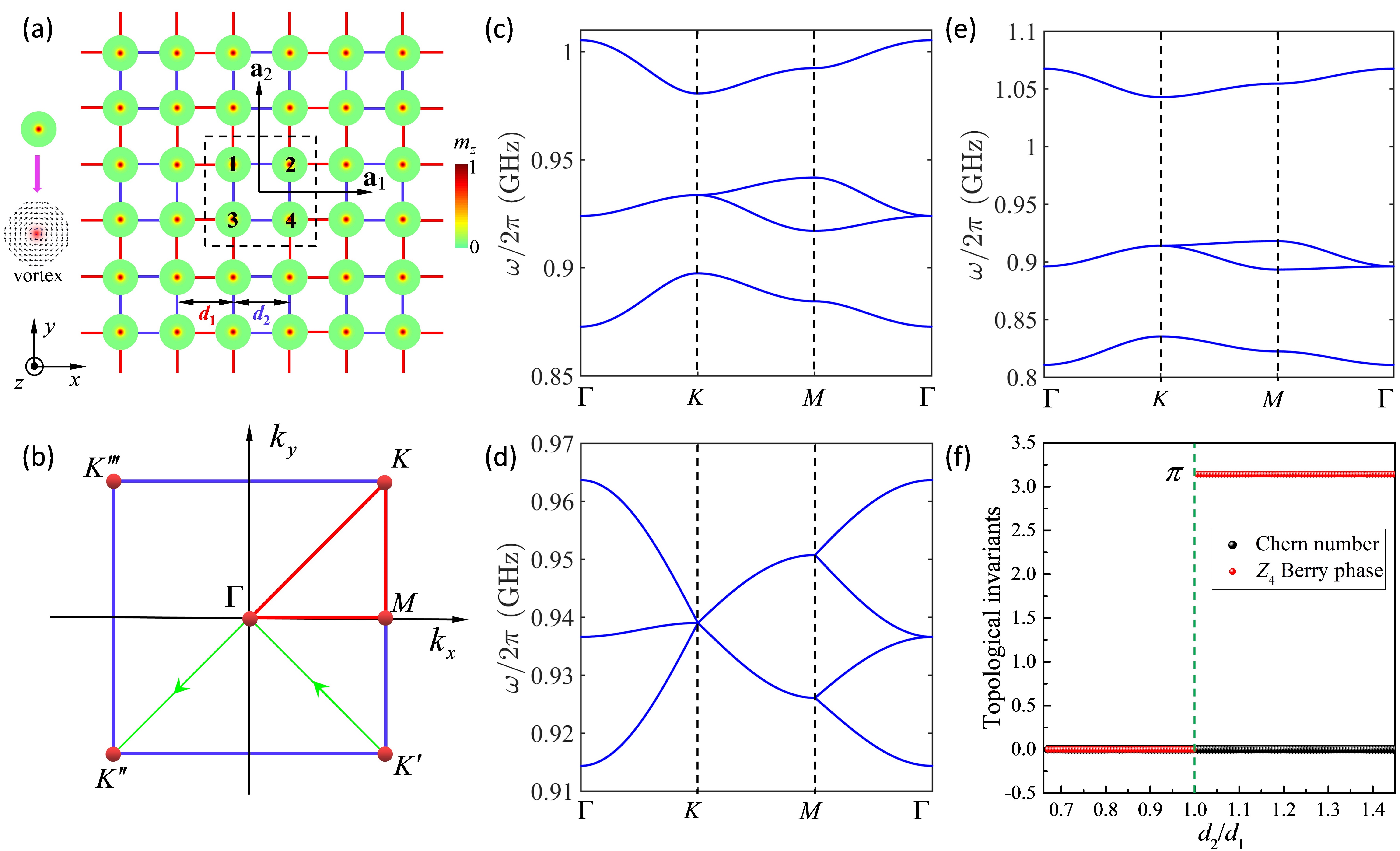}
\par\end{centering}
\caption{(a) Illustration of the breathing square lattice of magnetic vortices, with $d_{1}$ and $d_{2}$ denoting the alternating lengths of intercellular and intracellular bonds, respectively. The dashed black rectangle is the unit cell for calculating the band structure and topological invariants, with $\mathbf{a}_1$ and $\mathbf{a}_2$ denoting the basis vectors. Inset shows the micromagnetic structure of the vortex, with thickness $w=10$ nm and radius $r=50$ nm. (b) The first Brillouin zone, with the high-symmetry points $\Gamma$, $K$, and $M$ locating at $(k_{x},k_{y})=(0,0)$, $(\frac{\pi}{a},\frac{\pi}{a})$, and $(\frac{\pi}{a},0)$, respectively. The (lowest-four) band structures along the path $\Gamma$-$K$-$M$-$\Gamma$ for different geometric parameters: $d_{1}=3.6r, d_{2}=2.4r$ (c), $d_{1}=d_{2}=3.6r$ (d), and $d_{1}=2.08r, d_{2}=3.6r$ (e). (f) Dependence of Chern number and $\mathbb{Z}_{4}$ Berry phase on the ratio $d_{2}/d_{1}$ with $d_{1}$ being fixed to $3r$.}
\label{Figure1}
\end{figure*}

In this paper, we present both analytical and numerical studies of the collective dynamics of dipolar-coupled magnetic vortices in a two-dimensional breathing square lattice. By solving the generalized Thiele's equation, we obtain the band structure of the collective vortex gyrations. We derive the full phase diagram of the system by evaluating the topological invariants Chern number and $\mathbb{Z}_{4}$ Berry phase. Two different phases are identified: the trivial insulating phase and the second-order topological phase with the phase transition crossing the border $d_{2}/d_{1}=1$. Here $d_{1}$ and $d_{2}$ are the alternating intercellular and intracellular bond lengths, respectively, as shown in Fig. \ref{Figure1}(a). We discover three corner states possessing very different frequencies in a finite lattice when the system is in the HOTI phase. Full micromagnetic simulations are performed to verify theoretical predictions with an excellent agreement. Finally, we design a vortex-based imaging device, as a demonstration of the realistic application of the HOTI state.

The paper is organized as follows: Model and method are presented in Sec. \ref{section2}. Band structure and associated topological invariants (Chern number and $\mathbb{Z}_{4}$ Berry phase) are evaluated. Section \ref{section3} gives the main results, including theoretical computation of the corner states (Sec. \ref{section3}A), micromagnetic simulations (Sec. \ref{section3}B), and HOTI imaging device design (Sec. \ref{section3}C). Discussion and conclusion are drawn in Sec. \ref{section4}. Model parameters and theory of generalized chiral symmetry are given in the Appendixes.

\section{MODEL AND METHOD} \label{section2}

We consider a breathing square lattice of magnetic nanodisks with vortex states, see Fig. \ref{Figure1}(a). The generalized Thiele's equation \cite{Linpj2019,Li2019,Li2018PRB} is adopted to describe the collective dynamics of the vortex lattice:
\begin{equation}\label{Eq1}
  \mathcal {G}_{3}\hat{z}\times\frac{d^{3}\textbf{U}_{j}}{dt^{3}}-\mathcal {M}\frac{d^{2}\textbf{U}_{j}}{dt^{2}}+\mathcal {G}\hat{z}\times \frac{d\textbf{U}_{j}}{dt}+\textbf{F}_{j}=0,
\end{equation}
where $\mathbf{U}_{j}= \mathbf R_{j} - \mathbf R_{j}^{0}$ is the displacement of the $j$-th vortex core from the equilibrium position $\mathbf R_{j}^{0}$; $\mathcal {G}= -4\pi$$Qw M_{s}$/$\gamma$ is the gyroscopic constant, with $Q=\frac{1}{4\pi}\int \!\!\! \int{dxdy\mathbf{m}\cdot(\frac {\partial \mathbf{m}}{\partial {x} } \times \frac {\partial \mathbf{m}}{\partial y } )}$ being the topological charge [$Q=+1/2$ for our vortex configuration shown in Fig. \ref{Figure1}(a)]; $\mathbf {m}$ is the unit vector along the local magnetization direction; $w$ is the thickness of the nanodisk; $M_{s}$ is the saturation magnetization; $\gamma$ is the gyromagnetic ratio; $\mathcal {M}$ is the effective mass of the magnetic vortex \cite{Makhfudz2012,Yang2018OE,Buttner2015}; and $\mathcal {G}_{3}$ is the third-order non-Newtonian gyroscopic coefficient \cite{Mertens1997,Ivanov2010,Cherepov2012}. The conservative force can be expressed as $\textbf{F}_{j}=-\partial \mathcal {W} / \partial \mathbf U_{j}$. Here $\mathcal{W}$ is the total potential energy due to both the confinement from a single disk and the interaction between nearest neighbor disks: $\mathcal {W}=\sum_{j}\mathcal {K}\textbf{U}_{j}^{2}/2+\sum_{j\neq k}U_{jk}/2$ with $U_{jk}=\mathcal {I}_{\parallel}U_{j}^{\parallel}U_{k}^{\parallel}-\mathcal {I}_{\perp}U_{j}^{\perp}U_{k}^{\perp}$ \cite{Kim2017,Shibata2003,Shibata2004}. Here, $\mathcal {K}$ is the spring constant, $\mathcal {I}_{\parallel}$ and $\mathcal {I}_{\perp}$ are the longitudinal and transverse coupling constants, respectively. By imposing $\mathbf{U}_{j}=(u_{j},v_{j})$ and defining $\psi_{j}=u_{j}+i v_{j}$, we have
\begin{equation}\label{Eq2}
   \hat{\mathcal {D}}\psi_{j}=\omega_{K}\psi_{j}+\sum_{k\in\langle j\rangle}(\zeta_{l}\psi_{k}+\xi_{l} e^{i2\theta_{jk}}\psi^{*}_{k}),
\end{equation}
where the differential operator $\hat{\mathcal {D}}=i\omega_{3}\frac{d^{3}}{dt^{3}}-\omega_{M}\frac{d^{2}}{dt^{2}}-i\frac{d}{dt}$, $\omega_{3}=\mathcal {G}_{3}/\mathcal {|G|}$, $\omega_{M}=\mathcal {M}/\mathcal {|G|} $, $\omega_{K}=\mathcal {K}/\mathcal {|G|} $, $\zeta_{l}=(\mathcal {I}_{\parallel, l}-\mathcal {I}_{\perp, l})/2\mathcal {|G|} $, and $\xi_{l}=(\mathcal {I}_{\parallel, l}+\mathcal {I}_{\perp, l})/2\mathcal {|G|}$, in which $l=1$ ($l=2$) represents the intercellular (intracellular) connection. $\theta_{jk}$ is the angle of the direction $\hat{e}_{jk}$ from the $x$-axis, where $\hat{e}_{jk}=(\mathbf{R}_{k}^{0}-\mathbf{R}_{j}^{0})/|\mathbf{R}_{k}^{0}-\mathbf{R}_{j}^{0}|$ and $\langle j\rangle$ is the set of nearest intercellular and intracellular neighbors of $j$. Parameters $\mathcal {G}_{3}$, $\mathcal {M}$, and $\mathcal {K}$ can be determined from micromagnetic simulations (see Appendix A for details). The analytical expression of $\mathcal {I}_{\parallel}$ and $\mathcal {I}_{\perp}$ on the distance $d$ between vortices has been obtained in a simplified two-nanodisk system \cite{Linpj2019,Li2019}.

We then recast the complex variable $ \psi_{j}$ as
\begin{equation}\label{Eq3}
  \psi_{j}=\chi_{j}(t)\exp(-i\omega_{0}t)+\eta_{j}(t)\exp(i\omega_{0}t),
\end{equation}
with $\omega_{0}$ is the eigenfrequency of a single vortex gyration. For vortex gyrations with $Q=+1/2$, one can justify $|\chi_{j}|\ll|\eta_{j}|$. By substituting \eqref{Eq3} into \eqref{Eq2}, we have
\begin{widetext}
\begin{equation}\label{Eq4}
  \begin{aligned}
 \hat{\mathcal {D}}\psi_{j}=(\omega_{K}-\frac{\xi^{2}_{1}+\xi^{2}_{2}}{\bar{\omega}_{K}})\psi_{j}+\zeta_{1}\sum_{k\in\langle j_{1}\rangle}\psi_{k}+\zeta_{2}\sum_{k\in\langle j_{2}\rangle}\psi_{k}
   -\frac{\xi_{1}\xi_{2}}{2\bar{\omega}_{K}}\sum_{s\in\langle\langle j_{1}\rangle\rangle}e^{i2\bar{\theta}_{js}}\psi_{s}-\frac{\xi^{2}_{2}}{2\bar{\omega}_{K}}\sum_{s\in\langle\langle j_{2}\rangle\rangle}e^{i2\bar{\theta}_{js}}\psi_{s}-\frac{\xi^{2}_{1}}{2\bar{\omega}_{K}}\sum_{s\in\langle\langle j_{3}\rangle\rangle}e^{i2\bar{\theta}_{js}}\psi_{s},
  \end{aligned}
\end{equation}
\end{widetext}
where $\bar{\omega}_{K}=\omega_{K}-\omega_{0}^{2}\omega_{M}$, $\bar{\theta}_{js}=\theta_{jk}-\theta_{ks}$ is the relative angle from the bond $k\rightarrow s$ to the bond $j\rightarrow k$ with $k$ between $j$ and $s$, and $\langle j_{1}\rangle$ and $\langle j_{2} \rangle$ ($\langle\langle j_{1}\rangle\rangle$, $\langle\langle j_{2}\rangle\rangle$, and $\langle\langle j_{3}\rangle\rangle$) are the set of nearest (next-nearest) intercellular and intracellular neighbors of $j$, respectively.

For an infinite lattice, the dashed black rectangle indicates the unit cell, as shown in Fig. \ref{Figure1}(a). $\textbf{a}_{1}=a\hat{x}$ and $\textbf{a}_{2}=a\hat{y}$ are two basis vectors, with $a=d_{1}+d_{2}$. We then obtain the matrix form of the Hamiltonian in momentum space by considering a plane wave expansion of $\psi_{j}=\phi_{j}\exp(i\omega t)\exp\big[i(n\mathbf{k}\cdot\textbf{a}_{1}+m\textbf{k}\cdot\textbf{a}_{2})\big]$, where $\textbf{k}$ is the wave vector, $n$ and $m$ are two integers:

\begin{widetext}
\begin{equation}\label{Eq5}
 \mathcal {H}=\left(
 \begin{matrix}
   Q_{0} & \zeta_{2}+\zeta_{1}\exp(-i\textbf{k}\cdot\textbf{a}_{1}) & \zeta_{2}+\zeta_{1}\exp(i\textbf{k}\cdot\textbf{a}_{2}) & Q_{1} \\
   \zeta_{2}+\zeta_{1}\exp(i\textbf{k}\cdot\textbf{a}_{1}) & Q_{0} & Q_{2} & \zeta_{2}+\zeta_{1}\exp(i\textbf{k}\cdot\textbf{a}_{2}) \\
   \zeta_{2}+\zeta_{1}\exp(-i\textbf{k}\cdot\textbf{a}_{2}) & Q_{2}^{*} & Q_{0}& \zeta_{2}+\zeta_{1}\exp(-i\textbf{k}\cdot\textbf{a}_{1})\\
   Q_{1}^{*} & \zeta_{2}+\zeta_{1}\exp(-i\textbf{k}\cdot\textbf{a}_{2})& \zeta_{2}+\zeta_{1}\exp(i\textbf{k}\cdot\textbf{a}_{1})& Q_{0}
  \end{matrix}
  \right),
\end{equation}
with elements explicitly expressed as
\begin{equation}\label{Eq6}
\begin{aligned}
Q_{0}&=\omega_{K}-\frac{\xi_{1}^{2}+\xi_{2}^{2}}{\bar{\omega}_{K}}-\frac{\xi_{1}\xi_{2}}{\bar{\omega}_{K}}[\cos(\textbf{k}\cdot\textbf{a}_{1})+\cos(\textbf{k}\cdot\textbf{a}_{2})], \\
Q_{1}&=\frac{\xi_{1}\xi_{2}}{\bar{\omega}_{K}}[\exp(i\textbf{k}\cdot\textbf{a}_{2})+\exp(-i\textbf{k}\cdot\textbf{a}_{1})]+\frac{\xi^{2}_{1}}{\bar{\omega}_{K}}\exp[i\textbf{k}\cdot(\textbf{a}_{2}-\textbf{a}_{1})]+\frac{\xi^{2}_{2}}{\bar{\omega}_{K}}, \\
Q_{2}&=\frac{\xi_{1}\xi_{2}}{\bar{\omega}_{K}}[\exp(i\textbf{k}\cdot\textbf{a}_{2})+\exp(i\textbf{k}\cdot\textbf{a}_{1})]+\frac{\xi^{2}_{1}}{\bar{\omega}_{K}}\exp[i\textbf{k}\cdot(\textbf{a}_{2}+\textbf{a}_{1})]+\frac{\xi^{2}_{2}}{\bar{\omega}_{K}}.
\end{aligned}
\end{equation}
\end{widetext}

Topological invariant Chern number can be used to judge whether the system is in the first-order TI phase \cite{Avron1983,WangPRB2017}:
\begin{equation}\label{Eq7}
   \mathcal{C}=\frac{i}{2\pi}\int\!\!\!\int_{\text{BZ}}dk_{x}dk_{y}\text{Tr}\big[P( \frac{\partial P}{\partial k_{x}}\frac{\partial P}{\partial k_{y}}- \frac{\partial P}{\partial k_{y}}\frac{\partial P}{\partial k_{x}})\big],
\end{equation}
where $P$ is the projection matrix $P(\textbf{k})=\phi (\textbf{k})\phi (\textbf{k})^{\dag}$, with $\phi (\textbf{k})$ being the normalized eigenstate (column vector) of \eqref{Eq5}, and the integral is over the first Brillouin zone (BZ). However, to determine whether the system allows the HOTI phase, a different topological invariant should be considered.

In the presence of four-fold rotational ($C_{4}$) symmetry, topological invariant $\mathbb{Z}_{4}$ Berry phase is a powerful tool to characterize the HOTI state \cite{Li2019}:
\begin{equation}\label{Eq8}
 \mathcal{\theta}=\int_{L_{1}}\text{Tr}[\textbf{A}(\textbf{k})]\cdot d\textbf{k}\ \  (\text{mod}\ 2\pi),
\end{equation}
where $\textbf{A}(\textbf{k})$ is the Berry connection:
\begin{equation}\label{Eq9}
 \textbf{A}(\textbf{k})=i\Psi^{\dag}(\textbf{k})\frac{\partial}{\partial\textbf{k}}\Psi(\textbf{k}).
\end{equation}
Here, $\Psi(\textbf{k})=[\phi_{1}(\textbf{k})$,$\phi_{2}(\textbf{k})$,$\phi_{3}(\textbf{k})]$ is the 4 $\times$ 3 matrix composed of the eigenvectors of Eq. \eqref{Eq5} for the lowest three bands [the corner states exist between the 3rd and 4th bands, see Fig. \ref{Figure2}(a) and Fig. \ref{Figure3}(a)]. $L_{1}$ is an integral path in momentum space $K^{\prime}\rightarrow \Gamma\rightarrow K^{\prime\prime}$; see the green line segment in Fig. \ref{Figure1}(b). Here, we use the Wilson-loop approach to evaluate the Berry phase $\theta$ so that the difficulty of the gauge choice can be avoided \cite{Benalcazar2017,Bernevig2017}. Besides, because of the $C_{4}$ symmetry, the four high-symmetry points $K$, $K^{\prime}$, $K^{\prime\prime}$, and $K^{\prime\prime\prime}$ in the first BZ are equivalent [see Fig. \ref{Figure1}(b)]. Therefore, there are other three equivalent integral paths ($L_{2}: K^{\prime\prime}\rightarrow \Gamma\rightarrow K^{\prime\prime\prime}$, $L_{3}: K^{\prime\prime\prime}\rightarrow \Gamma\rightarrow K$, $L_{4}: K\rightarrow \Gamma\rightarrow K^{\prime}$) giving rise to identical $\theta$. It is also obvious that the integral along the path $L_{1}+L_{2}+L_{3}+L_{4}$ is zero. Thus, the $\mathbb{Z}_{4}$ Berry phase must be quantized as $\theta=\frac{2n\pi}{4}\ $ $(n=0,1,2,3)$. Therefore, by simultaneously quantifying the Chern number and the $\mathbb{Z}_{4}$ Berry phase, we can accurately characterize the emerging topological phases and their transition.

Of particular interest are the corner states that are related to the symmetry of system Hamiltonian. One can show that the emergence of topological corner states is protected by the generalized chiral symmetry of the quadripartite lattice. The detailed proof is presented in Appendix B.

\section{Second-order magnetic topological insulator} \label{section3}
\subsection{Theoretical results}
\begin{figure}[ptbh]
\begin{centering}
\includegraphics[width=0.48\textwidth]{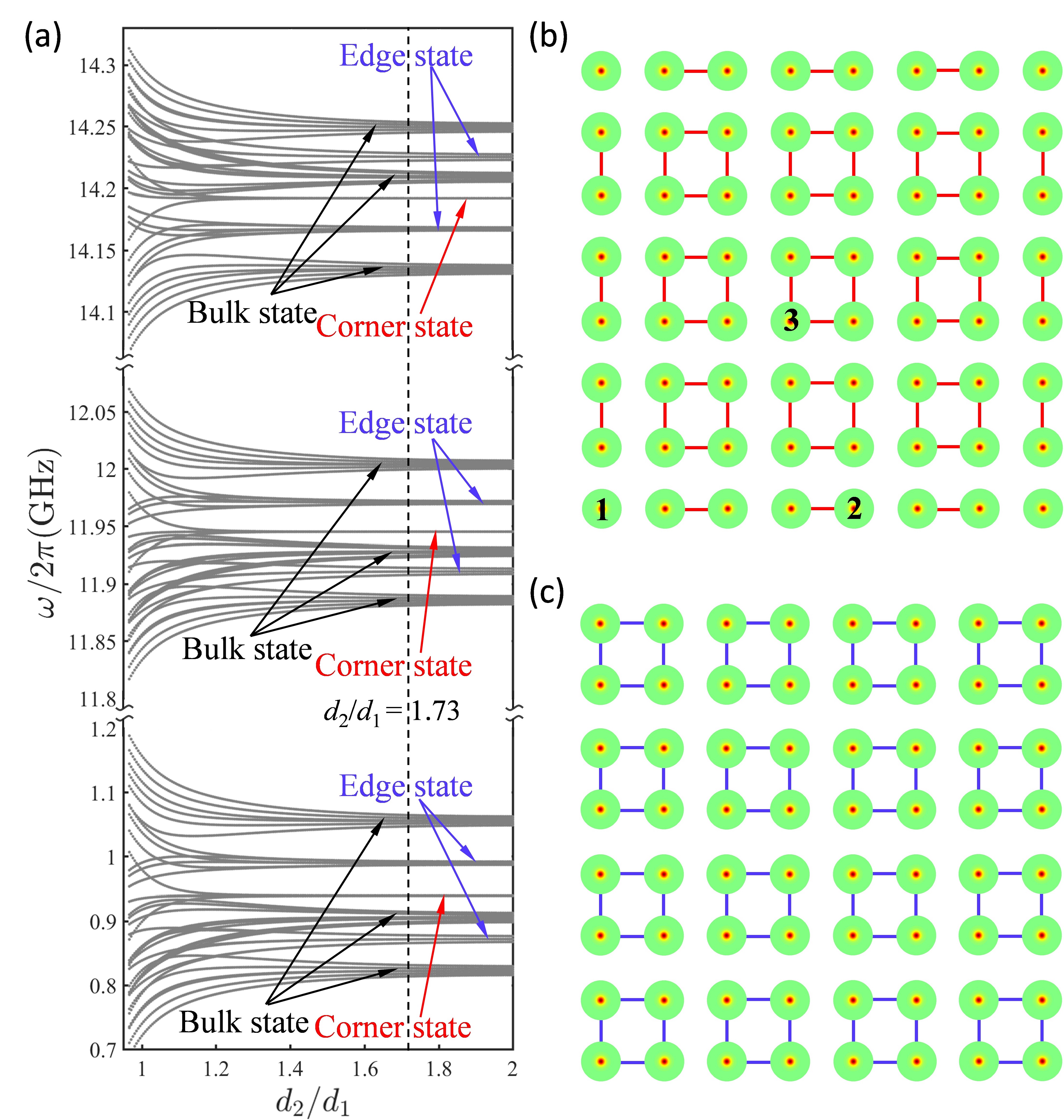}
\par\end{centering}
\caption{(a) Eigenfrequencies of collective vortex gyrations under different ratios $d_{2}/d_{1}$ in a finite lattice of the size $(3d_{1}+4d_{2})\times(3d_{1}+4d_{2})$. We set $d_{1}$ to $2.08r$ in the calculation. The schematic plot of the vortex lattice in (a) for two limit cases, $d_{2}\rightarrow\infty$ (b) and $d_{1}\rightarrow\infty$ (c).}
\label{Figure2}
\end{figure}
To investigate different phases allowed in the system, we calculate the bulk band structures with various geometric parameters ($d_{1}$ and $d_{2}$), as shown in Figs. \ref{Figure1}(c)-\ref{Figure1}(e) [we only plot the lowest four bands there]. For $d_{1}=d_{2}=3.6r$ [see Fig. \ref{Figure1}(d)], we find that all bands merge together, leading to a gapless band structure. However, when $d_{1}\neq d_{2}$, two gaps open and locates between 1st and 2nd bands, 3rd and 4th bands, respectively, see Figs. \ref{Figure1}(c) and \ref{Figure1}(e). Interestingly, the 2nd and 3rd bands are always merged no matter what values $d_{1}$ and $d_{2}$ take. To further distinguish whether these insulating phases are topologically protected, we examine simultaneously the topological invariants Chern number and $\mathbb{Z}_{4}$ Berry phase.

Figure \ref{Figure1}(f) plots the dependence of the Chern number $\mathcal{C}$ and the $\mathbb{Z}_{4}$ Berry phase $\mathcal{\theta}$ on the ratio $d_{2}/d_{1}$ with $d_{1}$ fixed to $3r$. In the calculations, the material parameters of Py \cite{Yoo2012,Velten2017} are adopted. We can clearly see that the $\mathbb{Z}_{4}$ Berry phase is quantized to 0 when $d_{2}/d_{1}<1$ and to $\pi$ otherwise, showing that $d_{2}/d_{1}=1$ is the phase transition point separating the trivial and topological phases. Furthermore, the Chern number vanishes for all ratios $d_{2}/d_{1}$, indicating that the system has no first-order TI phase. Therefore, we conclude that the system is in the HOTI phase when $d_{2}/d_{1}>1$, and in the trivial phase when $d_{2}/d_{1}<1$. It is worth noting that this conclusion holds independent of the $d_{1}$ value we choose.

To further confirm the existence of corner states in our system, we calculate the eigenfrequencies of collective vortex gyration as a function of $d_{2}/d_{1}$ in a finite square-shaped lattice [see Figs. \ref{Figure2}(b) and \ref{Figure2}(c)]. Numerical results are presented in Fig. \ref{Figure2}(a). By analyzing the spatial distribution of the eigenfunctions, we identify three different phases: bulk state, edge state, and corner state, marked by black, blue, and red arrows in Fig. \ref{Figure2}(a), respectively. The frequencies of three corner states are equal to those of a single vortex gyration (see Appendix A). To provide an intuitive understanding why these corner states only appear in the special parameter region ($d_{2}/d_{1}>1$), we plot the configuration of the lattice in the zero-correlation length limit, as shown in Fig. \ref{Figure2}(b) [Fig. \ref{Figure2}(c)] for $d_{2}\rightarrow\infty$ ($d_{1}\rightarrow\infty$). On the one hand, from the phase diagram [Fig. \ref{Figure1}(f)], we can infer that the configuration shown in Fig. \ref{Figure2}(b) is in the HOTI phase. In such a case, we clearly identify four isolated vortices at the corners of the lattice. One can thus observe localized corner states. On the other hand, in the limit $d_{1}\rightarrow\infty$ [see Fig. \ref{Figure2}(c)], there are no uncoupled vortices, thus no corner states. The system is therefore in the trivial phase.
\begin{figure}[ptbh]
\begin{centering}
\includegraphics[width=0.48\textwidth]{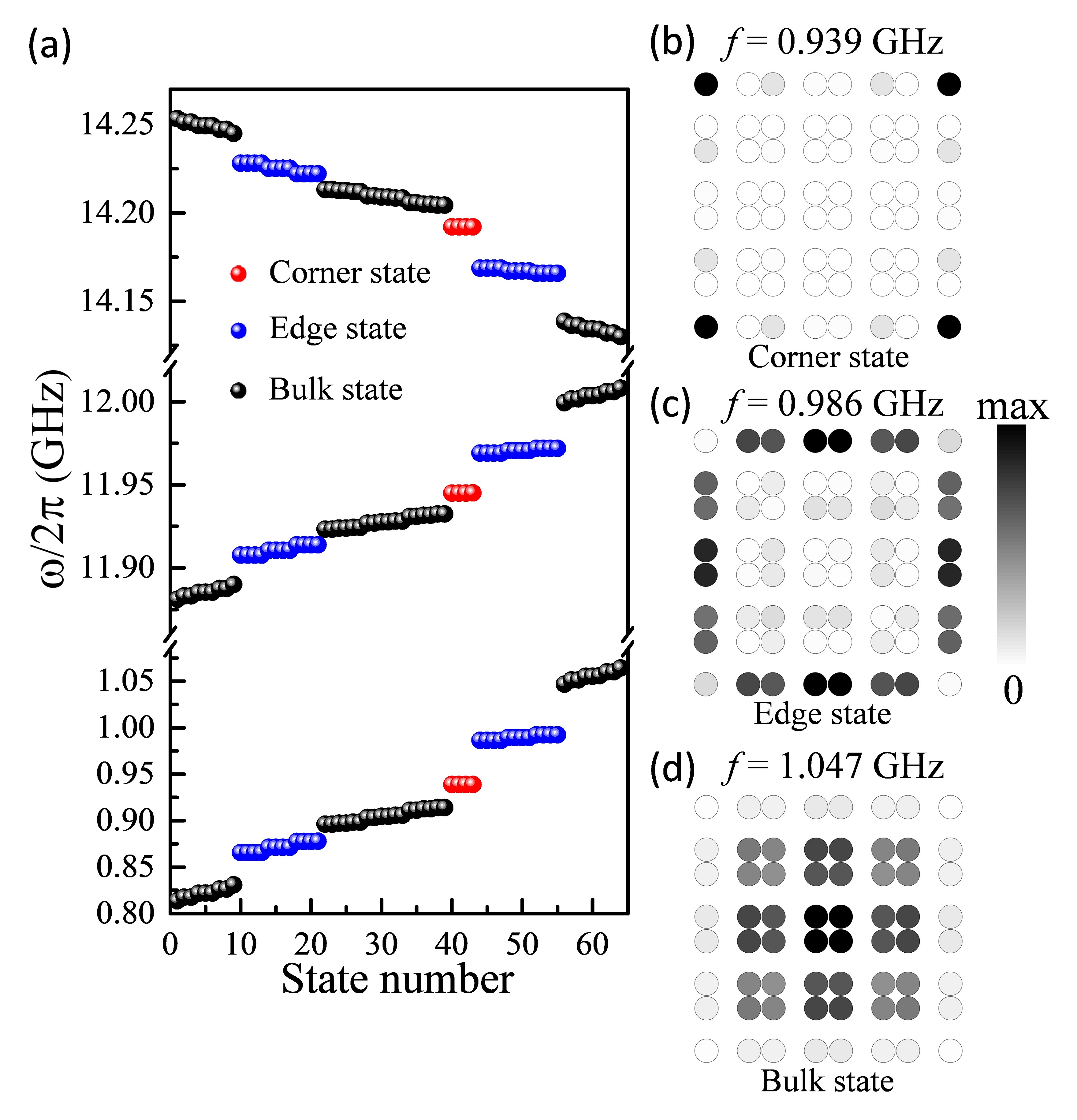}
\par\end{centering}
\caption{(a) Eigenfrequencies of square-shape vortex lattice with $d_{1}=2.08r$ and $d_{2}=3.6r$. The spatial distribution of vortex gyrations for the corner (b), edge (c), and bulk (d) states with the frequency 0.939 GHz, 0.986 GHz, and 1.047 GHz, respectively.}
\label{Figure3}
\end{figure}
\begin{figure}[ptbh]
\begin{centering}
\includegraphics[width=0.48\textwidth]{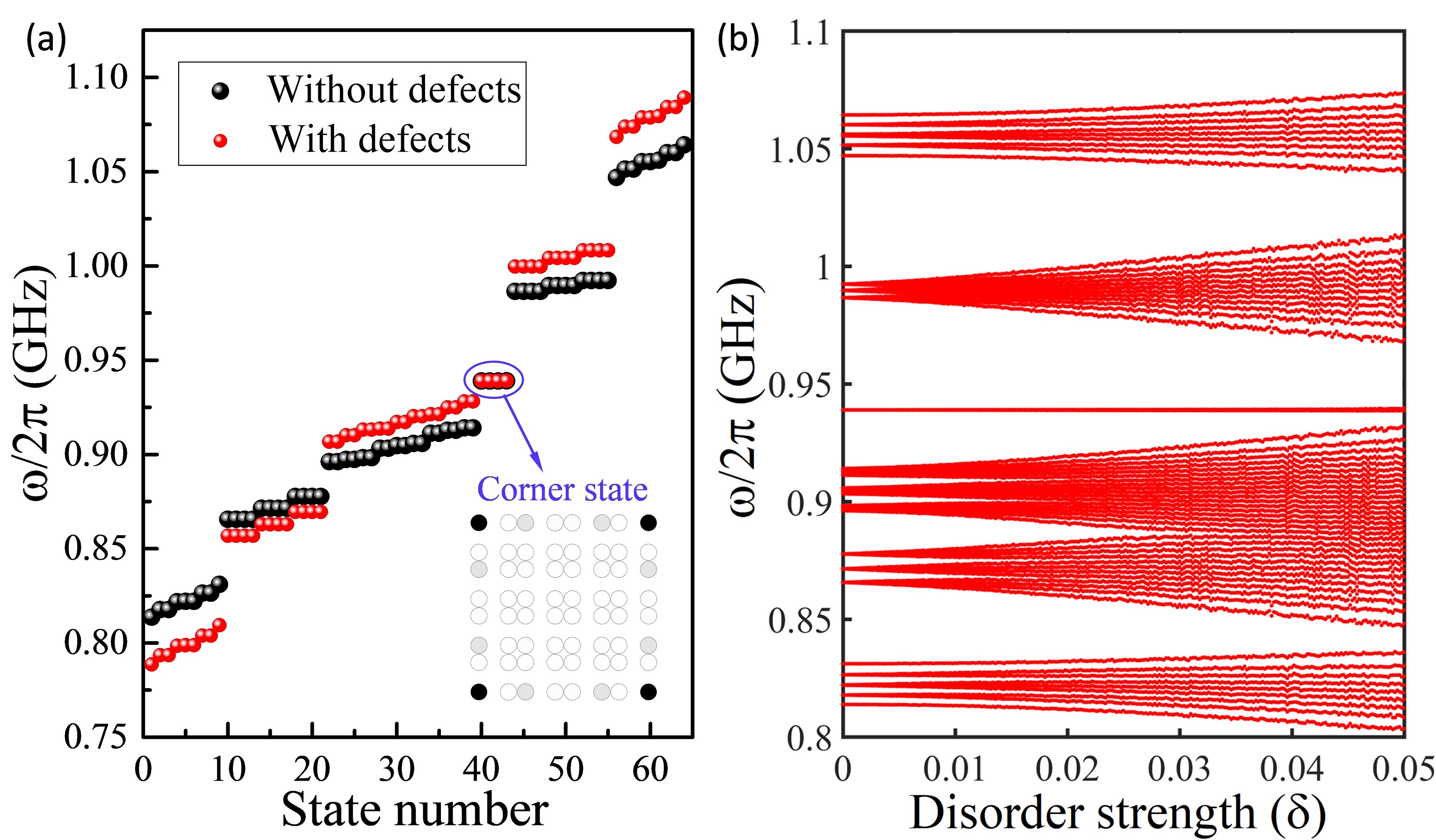}
\par\end{centering}
\caption{(a) Eigenfrequencies of the square-shaped vortex lattice in the absence of defects (black balls) an in the presence of defects (red balls). The blue circle indicates the topologically stable corner state with inset showing the corresponding spatial distribution of vortex gyrations in pristine lattice. (b) Eigenfrequencies of the square-shaped vortex lattice under different disorder strengths.}
\label{Figure4}
\end{figure}

To visualize the second-order corner states, we consider the square-shaped vortex lattice with $d_{1}=2.08r$ and $d_{2}=3.6r$, as indicated by the vertical dotted black line in Fig. \ref{Figure2}(a). Figure \ref{Figure3} shows the computed eigenfrequencies and eigenmodes. It is found that there exist three corner states with different frequencies (0.939 GHz, 11.945 GHz, and 14.192 GHz), represented by red balls in Fig. \ref{Figure3}(a). The spatial distribution of the corner state shows that its oscillation is highly localized at four corners [see Fig. \ref{Figure3}(b)]. We also identify the edge state, denoted by blue balls [see Fig. \ref{Figure3}(a)]. The spatial distribution of the edge state are confined along the lattice boundary, as shown in Fig. \ref{Figure3}(c). However, these edge states are Tamm-Shockley type \cite{Tamm1932,Shockley1939} and are not topologically protected because of the vanishing Chern number [see Fig. \ref{Figure1}(f)]. The spatial distribution of the bulk state [black balls in Fig. \ref{Figure3}(a)] is also plotted in Fig. \ref{Figure3}(d), where vortices at corners and edges do not participate in the oscillation.

To examine whether the corner states emerging in Fig. \ref{Figure3} have a topological nature, we calculate the eigenfrequencies of the square-shaped vortex lattice under defects and disorder, with numerical results presented in Figs. \ref{Figure4}(a) and \ref{Figure4}(b), respectively. Here, the defects are introduced by assuming a shift to the coupling parameters $\zeta$ and $\xi$, i.e., $\zeta\rightarrow1.2\zeta$ and $\xi\rightarrow0.8\xi$. The disorder is introduced by assuming the resonant frequency $\omega_{K}$ undergoing a random variation, i.e., $\omega_{K}\rightarrow\omega_{K}(1+\delta Z)$, where $\delta$ is the strength of disorder, $Z$ is a uniformly distributed random number between -1 to 1, which apply to all vortices except for those in corners. Numerical calculations have been averaged by 100 realizations. From Fig. \ref{Figure4}(a), we can clearly see that the frequency of the lowest corner state is perfectly pinned to 0.939 GHz, while the frequencies of edge and bulk states are significantly modified. Meanwhile, with the increasing of disorder strength, the corner state is very robust [see Fig. \ref{Figure4}(b)]. We therefore conclude that the corner states emerging in our system are topologically stable. It is worth mentioning that the other two high-frequency corner states around 11.945 GHz and 14.192 GHz share similar properties (not show here).
\begin{figure}[ptbh]
\begin{centering}
\includegraphics[width=0.48\textwidth]{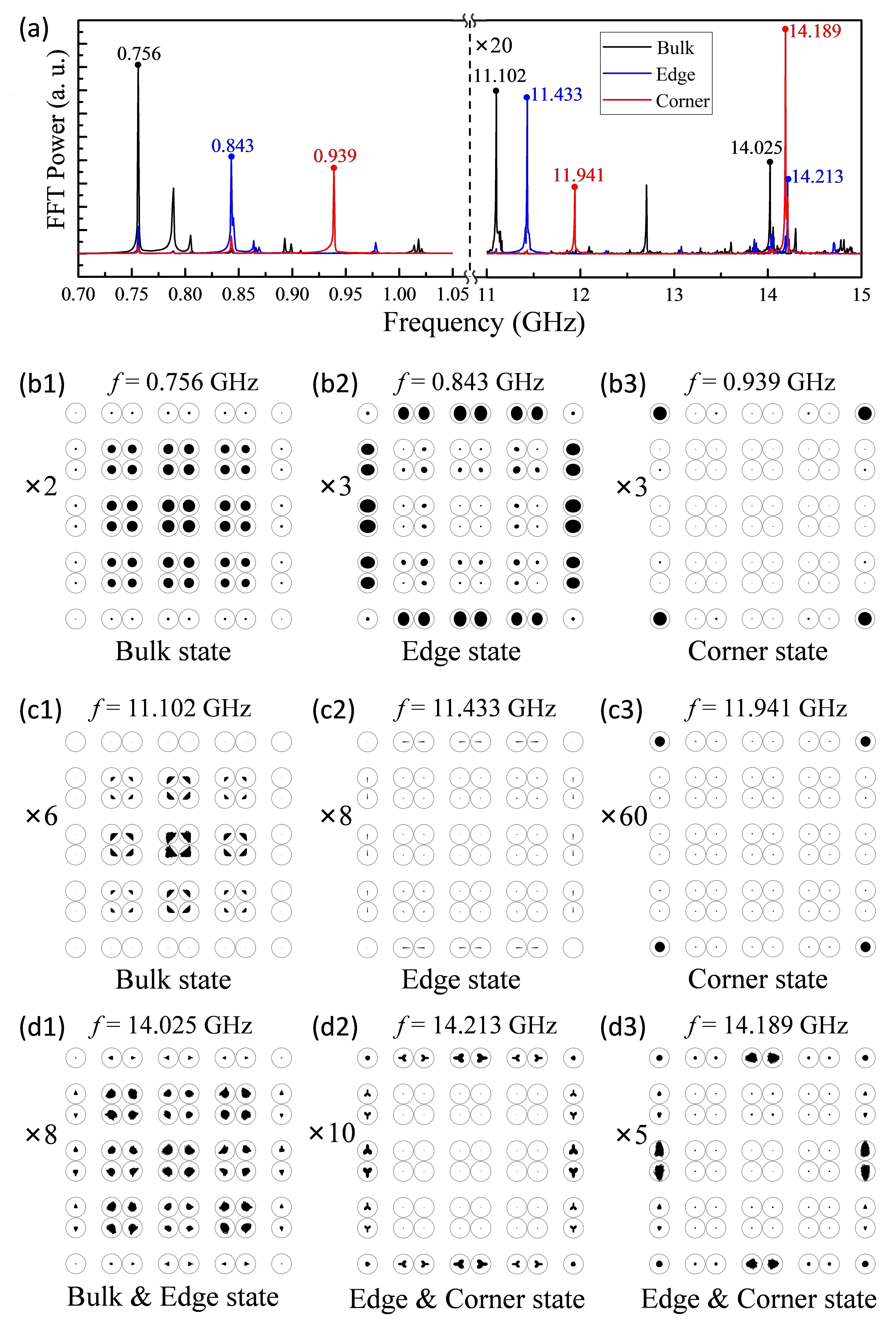}
\par\end{centering}
\caption{ (a) The temporal Fourier spectra of the vortex oscillations at different positions. (b1-d3) The spatial distribution of oscillation amplitude under the exciting field with different frequencies indicated in (a). Since the oscillation amplitudes of the vortex core are too small, we have magnified them by different times labeled in each figure.}
\label{Figure5}
\end{figure}

\subsection{Micromagnetic simulations}
To verify the theoretical predictions, we perform full micromagnetic simulations. All material parameters used in the simulations are the same as those for the theoretical calculations in Fig. \ref{Figure3}(a): the saturation magnetization $M_{s}=0.86\times10^{6}$ A\,m$^{-1}$, the exchange stiffness $A=1.3\times10^{-11}$ J\,m$^{-1}$, and the Gilbert damping constant $\alpha=10^{-4}$. The micromagnetic package MUMAX3 \cite{Vansteenkiste2014} is adopted to simulate the collective dynamics of the vortex lattice. The cellsize is set to $2\times2\times10 $ nm$^{3}$. To obtain the full spectrum, we apply a sinc-function magnetic field $H(t)=H_{0}\sin[2\pi$\emph{f}$(t-t_{0})]/[2\pi$\emph{f}$(t-t_{0})]$ along the $x$-direction with $H_{0}=10$ mT, $f=30$ GHz, and $t_{0}=1$ ns, over the whole system for 1 $\mu$s. The position of vortex cores $\textbf{R}_{j}=(R_{j,x}, R_{j,y}$) in all nanodisks are recorded every 20 ps. Here, $R_{j,x}=\frac {\int \!\!\! \int{x|m_{z}|^{2}dxdy}}{\int \!\!\! \int{|m_{z}|^{2}dxdy}}$ and $R_{j,y}=\frac {\int \!\!\! \int{y|m_{z}|^{2}dxdy}}{\int \!\!\! \int{|m_{z}|^{2}dxdy}}$, with the integral region confined in the $j$-th nanodisk.

We analyze the temporal Fourier spectra of the vortex oscillations at different positions [labeled by arabic numbers 1, 2, and 3, see Fig. \ref{Figure2}(b)]. Figure \ref{Figure5}(a) shows the spectra, with black, blue, and red curves representing the position of bulk, edge, and corner bands, respectively. We have magnified the FFT power by 20 times at the high frequency part ($>11$ GHz, as marked in the figure) for the reason is that the vortex-oscillation amplitudes are weak in this region. From Fig. \ref{Figure5}(a), we can clearly see that near the eigenfrequencies of a single vortex gyration (0.939 GHz and 11.941 GHz), the spectrum for the corner has two very strong peaks, which do not exist for edge and bulk bands. We thus infer that they are two corner states. Similarly, we identify the frequency range supporting the bulk and edge states, around 0.756 GHz (11.102 GHz) and 0.843 GHz (11.433 GHz), respectively. Interestingly, for the 14.189 GHz peak, although the spectrum in the corner has a strong peak, the oscillation amplitude at the edge is sizable as well, which indicates a strong coupling between edge and corner oscillations. Similar mode hybridization occurs at 14.025 GHz and 14.213 GHz, too. To visualize the spatial distribution of vortex oscillations for different modes mentioned above, we stimulate the dynamics of vortex lattice by applying a sinusoidal field $\textbf{h}(t)=h_0\sin(2\pi ft)\hat{x}$ with $h_{0}=0.1$ mT for 100 ns. Then we plot the spatial distribution of oscillation amplitude for different frequencies in Figs. \ref{Figure5}(b1)-(d3). One can distinguish the bulk states [Figs. \ref{Figure5}(b1) and \ref{Figure5}(c1)], edge states [Figs. \ref{Figure5}(b2) and \ref{Figure5}(c2)], and corner states [Figs. \ref{Figure5}(b3) and \ref{Figure5}(c3)]. The hybridized modes are observed as well: bulk \& edge state [Fig. \ref{Figure5}(d1)] and edge \& corner state [Figs. \ref{Figure5}(d2) and \ref{Figure5}(d3)]. It is worth noting that the mode hybridization results from the fact that the frequencies of these different states are so close, see Fig. \ref{Figure5}(a).

\subsection{HOTI imaging device}
\begin{figure}[ptbh]
\begin{centering}
\includegraphics[width=0.48\textwidth]{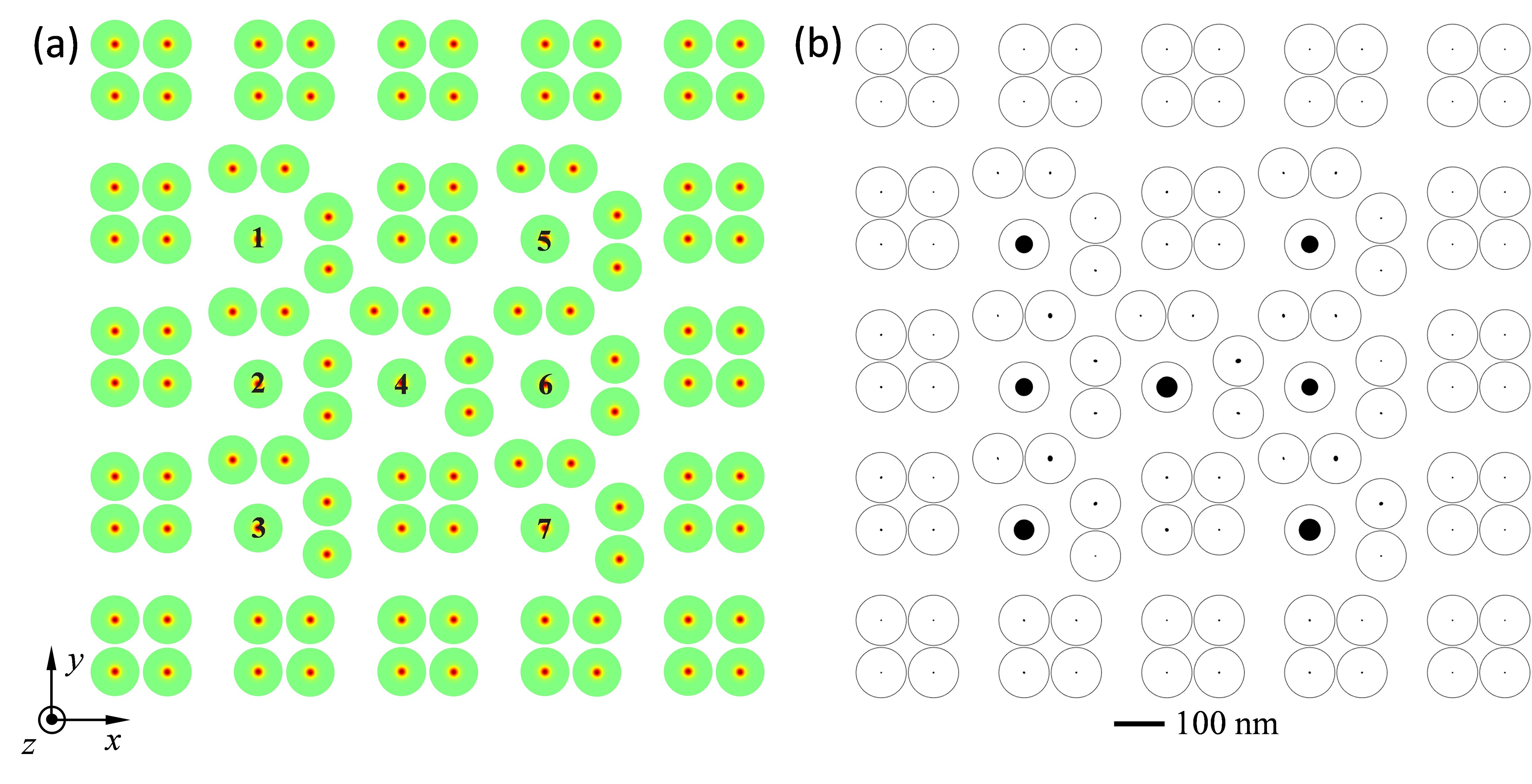}
\par\end{centering}
\caption{ (a) The schematic plot of the vortex lattice for ``H'' imaging. (b) Micromagnetic simulation of vortex gyrations with frequency $f=0.939$ GHz. The oscillation amplitudes of the vortex core have been magnified by 3 times.}
\label{Figure6}
\end{figure}

To demonstrate a practical application of topologically stable corner states, we design an imaging device based on the vortex lattice. The desired imaging ``H'' in the HOTI phase is surrounded by another vortex lattice in the trivial phase, as shown in Fig. \ref{Figure6}(a). The imaging points are marked by arabic numbers 1 to 7. Then we stimulate the collective dynamics of the whole system by applying a sinusoidal magnetic field with frequency $f=0.939$ GHz and amplitude $h_{0}=0.1$ mT along the $x$-direction for 80 ns. Figure \ref{Figure6}(b) plots the spatial distribution of the oscillation amplitudes, from which we can clearly see that only the vortices at the desired imaging points have sizable oscillations, while the other vortices do not participate in the imaging. We therefore demonstrate an ``H'' imaging device. We expect that other imaging shapes can be realized by similar method, too.

\section{DISCUSSION AND CONCLUSION} \label{section4}
To conclude, we have studied the collective dynamics of a breathing square lattice of magnetic vortices. The full phase diagram was obtained theoretically by computing the Chern number and $\mathbb{Z}_{4}$ Berry phase. Two different phases (the trivial and the second-order topological phases) were identified with the phase transition point at $d_{2}/d_{1}=1$. By including higher-order modifications in Thiele's equation, we predicted the existence of three corner states with different frequencies varying from sub GHz to tens of GHz. The emerging corner states are shown to be robust against moderate defects and disorder because of the topological protection from the generalized chiral symmetry in quadripartite lattices. Full micromagnetic simulations matched theoretical predictions with an excellent agreement. We finally designed a HOTI imaging device, as a demonstration of practical applications. From an experimental point of view, the artificial vortex lattices can be easily fabricated by electron-beam lithography \cite{Han2013,Behncke2015,Sun2013}, and the nanometer-scale vortex orbits can be tracked by using the ultrafast Lorentz microscopy technique in a time-resolved manner \cite{Moller}. We envision the existence of third-order topological states in three-dimensional breathing square lattice of magnetic vortices, which is an interesting issue for future study. In addition, the topological property of twisted bilayer of magnetic soliton lattice is also an appealing research topic.

\begin{acknowledgments}
\section*{ACKNOWLEDGMENTS}
We thank Zhenyu Wang for helpful discussions. This work was supported by the National Natural Science Foundation of China (NSFC) (Grants No. 11604041 and 11704060), the National Key Research Development Program under Contract No. 2016YFA0300801, and the National Thousand-Young-Talent Program of China. X. R. Wang was supported by Hong Kong RGC (Grants No. 16300117, 16301518, and 16301619). Z.-X. Li acknowledges the financial support of the China Postdoctoral Science Foundation
(Grant No. 2019M663461) and the NSFC Grant No. 11904048.
\end{acknowledgments}

\section*{APPENDIX A: MODEL PARAMETERS}
The non-Newtonian gyroscopic coefficient $G_{3}$, effective mass $M$, and spring constant $K$ are important parameters for evaluating the band structures. Here we determinate these parameters by considering the dynamics of a single vortex confined in a nanodisk. We start with the generalized Thiele's equation \eqref{Eq1}. In this case, the potential energy reads $\mathcal {W}=\mathcal {W}_{0}+\mathcal {K}\textbf{U}_{j}^{2}/2$. Assuming $\psi_{j}(t)=\psi_{j} e^{i\omega_0 t}$, Eq. \eqref{Eq1} can be simplified to
\begin{equation}\label{Eq10}
  \mathcal {G}_{3}\omega_{0}^{3}+\mathcal {M}\omega_{0}^{2}-\mathcal {G}\omega_{0}-\mathcal {K}=0.
\end{equation}
In our system, we have $\mathcal {G}=-3.0725\times10^{-13}$ J\,s\,rad$^{-1}$m$^{-2}$. Moreover, the three eigenfrequencies for a single vortex oscillation in a nanodisk are \cite{Linpj2019,Li2019}: $\omega_{0,1}/2\pi=+0.939$ GHz,  $\omega_{0,2}/2\pi=-11.945$ GHz, and $\omega_{0,3}/2\pi=+14.192$ GHz. Here, the sign $+$ ($-$) represents the vortex gyration direction to be anti-clockwise (clockwise). Solving Eq. \eqref{Eq10} with the three eigenfrequencies, we obtain $\mathcal{G}_{3}=-4.6488\times10^{-35}$J$\,$s$^{3}$rad$^{-3}$m$^{-2}$, $\mathcal{M}=9.3061\times10^{-25}$ kg, and $\mathcal{K}=1.8356\times10^{-3}$ J$\,$m$^{-2}$.

\section*{APPENDIX B: GENERALIZED CHIRAL SYMMETRY IN QUADRIPARTITE LATTICES}

Here, we prove that the emerging topological corner states in our system are protected by the generalized chiral symmetry for quadripartite lattice. First of all, because $\frac{\xi_{1}^{2}+\xi_{2}^{2}}{\bar{\omega}_{K}}+\frac{\xi_{1}\xi_{2}}{\bar{\omega}_{K}}[\cos(\textbf{k}\cdot\textbf{a}_{1})+\cos(\textbf{k}\cdot\textbf{a}_{2})]\ll\omega_{K}$, the diagonal element of $\mathcal {H}$ can be regarded as a constant $Q_{0}=\omega_{K}$, which is the ``zero-energy" of the Hamiltonian. Then we generalize the chiral symmetry for a unit cell containing four sites by defining
\begin{equation}\label{Eq11}
\begin{aligned}
 \Gamma_{4}^{-1}\mathcal {H}_{1}\Gamma_{4}&=\mathcal {H}_{2},\\
 \Gamma_{4}^{-1}\mathcal {H}_{2}\Gamma_{4}&=\mathcal {H}_{3},\\
 \Gamma_{4}^{-1}\mathcal {H}_{3}\Gamma_{4}&=\mathcal {H}_{4},\\
 \mathcal {H}_{1}+\mathcal {H}_{2}+\mathcal {H}_{3}&+\mathcal {H}_{4}=0,
 \end{aligned}
\end{equation}
with the chiral operator $\Gamma_{4}$ is a diagonal matrix to be determined, and $\mathcal{H}_{1}=\mathcal{H}-Q_{0}\text{I}$. By combining the last equation with the previous three in Eqs. \eqref{Eq11}, we have $\Gamma_{4}^{-1}\mathcal {H}_{4}\Gamma_{4} =\mathcal {H}_{1}$, indicating that $[\mathcal {H}_{1},\Gamma_{4}^{4}]=0$; so that $\Gamma_{4}^{4}=\text{I}$, via the reasoning completely analogous to the Su-Schrieffer-Heeger model \cite{SSH1979}. In addition, Hamiltonians $\mathcal {H}_{1,2,3,4}$ each have the same eigenvalues $\lambda_{1,2,3,4}$. The eigenvalues of $\Gamma_{4}$ are given by $1,\ \exp(2\pi i/4),\ \exp(\pi i)$, and $\exp(6\pi i/4)$. We thus have
\begin{equation}\label{Eq12}
 \Gamma_{4}=\left(
 \begin{matrix}
   1 & 0 & 0& 0 \\
   0 & e^{\frac{2\pi i}{4}} & 0& 0 \\
   0 & 0 & e^{\pi i}& 0\\
   0 & 0 & 0& e^{\frac{6\pi i}{4}}
  \end{matrix}
  \right),
\end{equation}
in suitable bases. By taking the trace of the fourth line from Eqs. \eqref{Eq11}, we find $\text{Tr}(\mathcal {H}_{1}+\mathcal {H}_{2}+\mathcal {H}_{3}+\mathcal {H}_{4})=4\text{Tr}(\mathcal {H}_{1})=0$, which means that the sum of the four eigenvalues vanishes $\sum_{i=1}^{4}\lambda_{i}=0$. Given an eigenstate $\phi_{j}$ that has support in only sublattice $j$, it will satisfy $\mathcal {H}_{1}\phi_{j}=\lambda\phi_{j}$ and $\Gamma_{4}\phi_{j}=\exp[2\pi i(j-1)/4]\phi_{j}$ with $j=1,2,3,4$. From these formulas and Eqs. \eqref{Eq11}, we find that $(\mathcal {H}_{1}+\mathcal {H}_{2}+\mathcal {H}_{3}+\mathcal {H}_{4})\phi_{j}=\sum_{i=1}^{4}\Gamma_{4}^{-(i-1)}\mathcal {H}_{1}\Gamma_{4}^{i-1}\phi_{j}=4\lambda\phi_{j}=0$, indicating $\lambda=0$ for any mode that has support in only one sublattice, i.e., zero-energy corner state.

\end{document}